\documentclass[journal]{IEEEtran}

\usepackage[table]{xcolor}
\usepackage{cite}
\usepackage{amsmath,amssymb,amsfonts}
\usepackage{graphicx}
\usepackage{textcomp}
\usepackage{xcolor}
\usepackage{tabularx}
\usepackage{pifont}
\usepackage{placeins}
\usepackage{mathtools}

\usepackage{subfig}
\usepackage{bbold}
\usepackage[short]{optidef}

\usepackage[ruled,vlined,linesnumbered]{algorithm2e}
\usepackage{algorithmicx}
\usepackage{lipsum}
\usepackage{setspace}
\newcommand{\subparagraph}{}

\DeclareMathOperator*{\argmax}{arg\,max}

\DeclareMathOperator*{\diag}{diag}

\begin{document}

\providecommand{\keywords}[1]
{
  \textbf{\textit{Keywords---}} #1
}

	\title{
	RIS-Assisted UAV for Timely Data Collection in IoT Networks
	}

\DeclarePairedDelimiter\abs{\lvert}{\rvert}%
\DeclarePairedDelimiter\norm{\lVert}{\rVert}%
	\newcommand{\INDSTATE}[1][1]{\State\hspace{#1\algorithmicindent}}

	\makeatletter
\newcommand{\multiline}[1]{%
  \begin{tabularx}{\dimexpr\linewidth-\ALG@thistlm}[t]{@{}X@{}}
    #1
  \end{tabularx}
}

\newcommand{\thickhat}[1]{\mathbf{\hat{\text{$#1$}}}}
\newcommand{\thickbar}[1]{\mathbf{\bar{\text{$#1$}}}}
\newcommand{\thicktilde}[1]{\mathbf{\tilde{\text{$#1$}}}}

\makeatother	
	
	\author{\IEEEauthorblockN{Ahmed Al-Hilo\IEEEauthorrefmark{1}, Moataz Samir\IEEEauthorrefmark{1}, Mohamed Elhattab\IEEEauthorrefmark{1}, Chadi Assi\IEEEauthorrefmark{1}, and Sanaa Sharafeddine\IEEEauthorrefmark{2}}\\
    \IEEEauthorblockA{\IEEEauthorrefmark{1}Concordia University,}
    \IEEEauthorblockA{\IEEEauthorrefmark{2}Lebanese American University}
	}

	\maketitle

\newcommand\blfootnote[1]{%
  \begingroup
  \renewcommand\thefootnote{}\footnote{#1}%
  \addtocounter{footnote}{-1}%
  \endgroup
}

\begin{abstract}
Intelligent Transportation Systems (ITS) are thriving thanks to a wide range of technological advances, namely 5G communications, Internet of Things, artificial intelligence and edge computing. Central to this is the wide deployment of smart sensing devices and accordingly the large amount of harvested information to be processed for timely decision making. Robust network access is, hence,  essential for offloading the collected data before a set deadline, beyond which the data loses its value.  
In environments where direct communication can be impaired by, for instance, blockages such as in urban cities, unmanned aerial vehicles (UAVs) can be considered as an alternative for providing and enhancing connectivity, particularly when IoT devices (IoTD) are constrained with their resources. Moreover, to conserve energy, IoTDs are assumed to alternate between their active and passive modes. This paper, therefore, considers a time-constrained data gathering problem from a network of sensing devices and with assistance from a UAV. A Reconfigurable Intelligent Surface (RIS) is deployed to further help  improve the connectivity and energy efficiency of the UAV, particularly when the multiple devices are served concurrently and experience different channel impairments. This integrated problem brings challenges related to the configuration of the phase shift elements of the RIS, the scheduling of IoTDs transmissions as well as the trajectory of the UAV.
%
First, the problem is formulated with the objective of maximizing the total number of served devices each during its activation period. Owing to its complexity and the incomplete knowledge about the environment, we leverage deep reinforcement learning in our solution; the UAV trajectory planning is modeled as a Markov Decision Process (MDP), and Proximal Policy Optimization (PPO) is invoked to solve it. Next, the RIS configuration is then handled via Block Coordinate Descent (BCD). Finally, extensive simulations are conducted to demonstrate the efficiency of our solution approach that, in many cases, outperforms other methods by more than 50\%. We also show that integrating a RIS with a UAV in IoT networks can notably improve the UAV energy efficiency (For example, UAV energy efficiency is increased by 5 times when using a RIS of 100 elements).
\end{abstract}


\section{Introduction}
\label{sec-introduction}
Future networks are expected to render services for a multitude of applications with heterogeneous requirements in terms of ultra low latency, high reliability, ultra high bandwidth and massive device connectivity. Applications ranging from Internet of Things for smart cities, intelligent transportation and industrial processes to other consumer- and entertainment-centric services such as virtual and augmented reality (VR/AR) as well as holographic tele-presence, are emerging at a fast pace and accelerating the development of revolutionary technologies to meet such stringent requirements. 
According to a recent study by Ericsson, the number of IoT devices (IoTD) will grow consistently by 27\% over the coming years with more than 4 billion cellular IoT connections expected in 2024 \cite{chung2020design}. IoTDs are, and will continue to be, embedded in every aspect of our world ranging from home appliances, street sensors and cameras to humans' bodies and connected vehicles. The Internet of Things paradigm notably contributes to realize intelligent transportation systems (ITS) and will play a major role in enabling future ITS services, such as fully, or semi, automated driving. To enable such services, the ITS applications will rely on harvesting sensing information from sensors and other IoT devices embedded along roads and on-board vehicles, as well as pedestrians. Such information will be offloaded to cloud applications running at the network edge for analytics and decision making. One impediment for this architecture is, however, the limited energy resources  of sensors and IoTDs, which makes their connection to the network a bottleneck\cite{biason2018access}. Ideally, high bandwidth connections are needed to transfer the sensory data (e.g., videos, images, etc.) at a high rate. One solution is to deploy a large number of small cells (ultra densification in the context of 5G), which are interconnected through backhaul links to a base station hosting the edge servers. This solution may not be cost efficient however.

\begin{figure}[t]
	\centerline{\includegraphics[scale=0.6]{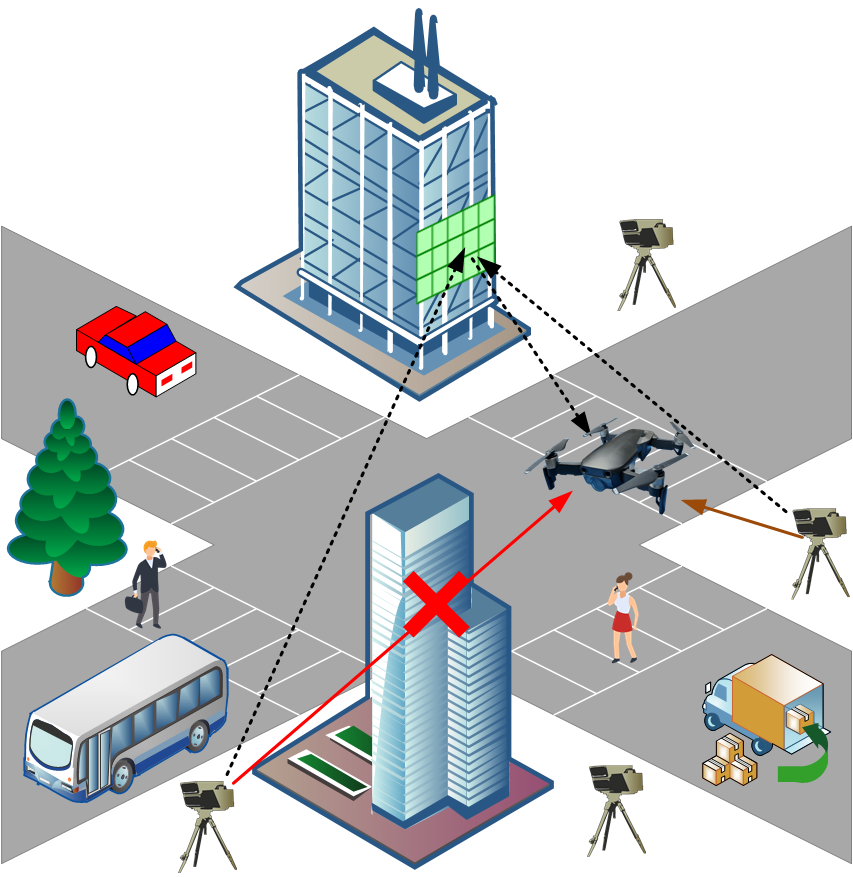}}
	\caption{System Model.}
	\label{fig:system_model_iot}
\end{figure}

Owing to their flexibility and ease of deployability, UAVs have been proposed by cellular operators to provide network extension for enhanced coverage, network connectivity during outages and on demand service during peak loads \cite{mozaffari2016unmanned}. UAVs are commonly proposed to also supplement IoT networks \cite{motlagh2017uav, abd2018average}. UAVs can communicate with a large number of IoT devices by adjusting their location to establish improved communication links. UAVs have been used particularly in the context of transportation systems to provide connectivity to ground vehicles \cite{samir2020trajectory} as well as to support IoT networks, e.g., to collect data \cite{samir2019uav} or enhance their transmissions \cite{huo2019distributed}. More specifically, in \cite{samir2019uav}, the authors studied the problem of optimizing the trajectory of a UAV to collect data in a timely manner. Given the disperse deployment of the IoT devices, the UAV may not be able to assist all such devices from one particular location, especially when devices are deployed in an urban environment where a direct line of sight is not always available. Recently, the concept of programming the propagation environment has emerged due to recent advances in meta surface materials \cite{basar2019wireless}. The idea behind this concept is to establish an indirect channel of communication between a pair of nodes when the direct channel is impaired by blockages and weak LoS. Reconfigurable Intelligent Surfaces (RIS) are such techniques and have been considered as a revolutionary technology which will advance the coverage of networks, in presence of blockages and impairments \cite{basar2019wireless, wu2019towards,Elhattab}. They are passive, cost effective and energy neutral, unlike their counterparts active relaying. We, in this paper, promote the integration between UAVs and terrestrial RIS elements and study the problem of collecting data from ground IoT devices (See Fig. \ref{fig:system_model_iot}). Although a UAV can sometimes establish LoS while flying at a high altitude, it is not always the case especially when considering concurrent communications with multiple terrestrial nodes (e.g. IoT devices) \cite{mozaffari2016unmanned}. Therefore, integrating RIS with UAV improves the quality of wireless channels and enhances UAV energy efficiency as will be shown in the numerical results. In this paper, we consider a UAV is deployed and its location is adapted to collect sensory data from active devices. Devices are assumed, to conserve energy, to switch between active and passive modes. Devices, for the purpose of facilitating some critical ITS services, are scattered and when a device is active it has sampled an information from a signal that it needs to send to the edge application. As mentioned, a UAV does not have a continuous LoS with all devices, and therefore, while it is in direct communication with some devices, other sensors may reach the UAV through an indirect path enabled by the RIS. We seek to concurrently optimize the RIS phase shift for its elements, the scheduling of IoT transmissions as well as the trajectory of the UAV.

Owing to the high complexity of the problem, we propose a solution based on Deep Reinforcement Learning (DRL) to cope with these challenges. DRL has been utilized in similar problems and showed to be both effective and efficient. Additionally, we suggest Block Coordinate Descent (BCD) to solve for the RIS phase-shift configuration. Indeed, serving IoTDs using UAV assisted by RIS, to the best of our knowledge, has never been investigated before. The contributions of this article can therefore be summarized as follows:

\begin{itemize}
    \item A UAV assisted with a RIS is leveraged to provide timely data gathering services from IoTDs. The UAV trajectory is optimized jointly with resource scheduling and RIS element configuration. This framework is formulated as an optimization problem aiming at maximizing two objectives; the total number of served IoTDs, and the UAV energy efficiency.
    
    \item Due to the high complexity of the formulated problem, which is a mixed-integer non-convex problem as well as the presence of unknown parameters (activation time of IoTDs), the formulated optimization problem is challenging and cannot be directly solved. In order to address this issue, we decompose the original optimization problem into two sub-problems. The first one is to determine the UAV mobility as well as IoTD scheduling. The second sub-problem, and with given UAV position and scheduled indices for the IoTDs, the passive beamforming problem, phase-shift matrix of the RIS, is solved.
    \item The first sub-problem is converted to Markov Decision Process (MDP) while tackling the massive action space incurred by UAV mobility in 2D space and IoTD scheduling. In addition, the state space is also defined to consider the UAV positioning and IoTDs' status while satisfying the Markov condition. After that, an agent based on proximal policy optimization (PPO) is developed to solve the MDP. Next, the second problem is solved via Block Coordinate Descent (BCD) where a low complexity method is designed to tune the RIS phase-shift to maximize the total transmission rate for all the IoTDs being served for each transmission.
    
    \item Finally, we carry out extensive simulation experiments to analyze the performance of our proposed solution in comparison with four baseline approaches. Furthermore, a study on the impact of RIS presence and size on the UAV energy efficiency is also conducted.
\end{itemize}

The organization of the rest of this paper is as follows. Section \ref{sec-literature-review} reviews the literature while showing the novel contributions of this article. Section \ref{sec-system-model} lays out the system model and presents our objective function. Section \ref{sec-solution-approach} describes the solution approach which is based on DRL and BCD. Section \ref{sec-numerical-results} studies the performance of the solution approach and signals out our observations. Finally, Section \ref{sec-conclution} wraps up the paper and provides some directions for future work.

\textit{Notations:} Vectors are denoted by bold-face italic letters. $\diag(x)$ denotes a diagonal matrix whose diagonal element is the corresponding element in $x$. $\mathbb{C}^{M \times N}$ denotes a complex matrix of $M \times N$. For any matrix $M$, $M^H$ and $M^T$ denote its conjugate transpose and transpose, respectively. $Pr(A \mid B)$ denotes the probability of event $A$ given event $B$.
\section{Literature Review}
\label{sec-literature-review}
Serving IoTDs has been studied widely in the literature. However, there are still gaps in terms of introducing RIS-empowered UAV communications to support IoT networks. In this section, we present some relevant publications in the area of data collection using UAVs, specifically for IoTDs, and RIS employment in wireless communication.

\subsection{UAVs for Data Collection}
In the literature, UAVs have been proposed to assist wireless communication for various purposes. The work of \cite{cai2020joint} suggests a framework for UAVs that provides UAV trajectory planning and resource allocation while considering the security aspects in the presence of eavesdroppers. \cite{lin2020dynamic} presents a formation of a number of UAVs that can collaborate to enhance the the service for a machine. The authors of \cite{shiri2020communication} suggest federated learning to control multiple UAVs with reduced overhead. In \cite{samir2019uav}, the trajectory of UAVs and their resources are jointly optimized to serve time-constrained IoTD. In \cite{samir2020online}, the authors propose UAV assisted IoTDs by optimizing their data freshness also known as Age of Information or AoI. Using DRL, the altitude of the UAV is changed to maintain a balance between the communication links of BS-UAV and IoTD-UAV. The authors of \cite{islambouli2019optimized} presents a UAV-based system where UAVs act as edge servers to offer computational resources for the IoTDs. The number of UAVs and their position in 3D space along with IoTD task allocation decisions are optimized jointly to provide service for the IoTD within a limited latency. In \cite{zhan2020energy}, an energy-constrained UAV is proposed to aid cellular communication by uploading data from base stations. To this end, the problem is formulated with the objective to maximize transmission throughput while considering resource scheduling, UAV trajectory, and energy budget. The authors in \cite{callegaro2021optimal} develop a framework that makes the UAV able to offload computational tasks received from IoTDs and others to a nearby facility. This work also addresses the problem of highly complex communication topology in urban environment and solves the concerned problem using DRL. The authors of \cite{liu2020resource} build a system model to collect IoTDs information using UAVs where the energy consumption of the IoTDs is minimized. This work also accounts for UAV 3D placement, resource allocation, wireless interference, and UAV altitude. In \cite{hu2019optimal}, a UAV is deployed to supply energy for ground nodes and optimizing the minimum energy provided. To do so, the trajectory of the UAV is optimized in 1D and the optimal solution is calculated.

\subsection{RIS-aided Wireless Communication}
Reflecting surfaces have gained momentum nowadays among the research communities owing to their notable benefits they bring to wireless communication systems. In \cite{yang2020performance}, the authors propose a relay UAV assisted by an RIS to receive signals from ground users. The RIS helps to improve the coverage and the overall performance of the network. The work of \cite{you2021enabling} discusses various scenarios for UAV integrated by RIS to improve wireless communications including data collection taking into account sensor nodes' transmit power. The authors of \cite{liu2020machine} develop a framework of UAV assisted by RIS using Non-orthogonal multiple access (NOMA) technique to enhance spectrum efficiency. To achieve this, the optimization problem is formulated to do UAV trajectory planning, resource allocation as well as RIS element configuration. Then, DRL is employed to solve the problem. The authors in \cite{wei2020sum} consider  frequency division multiple access for UAV assisted by RIS to optimize the max-sum rate. To do so, the UAV trajectory, RIS phase-shift, and resource allocation are jointly optimized while considering different quality-of-service for users. In \cite{shafique2020optimization}, a relaying framework for RIS integrated with UAV is investigated where spectral and energy efficiency is sought. Different modes are studied starting with UAV only and end up with UAV-RIS. The problem is formulated where RIS number of elements and UAV altitude are optimized.  In \cite{chen2020resource}, the authors studied resource allocation of RIS-aided vehicular communications where they aim to maximize vehicle to base station link quality while guaranteeing vehicle to vehicle communications. The authors of \cite{wang2020outage} provide analysis for outage probability in RIS-enabled vehicular networks. This paper derives an expression of outage probability showing that RIS can reduce the outage probability for vehicles in its vicinity. The analysis also proves that higher density roads increase outage probability since passing vehicles can block the communication links. In \cite{dampahalage2020intelligent}, the authors propose RIS-aided vehicular networks while considering two scenarios to estimate the channels. The first one is by assuming fixed channel estimation within a coherence time. While the second one neglects the small scale fading based on the fact that vehicular positions can be realised in advance. \cite{you2020channel} considers constraint discrete phase-shift RIS with two challenges; channel estimation and passive beamforming. The work in \cite{pan2020uav} proposes a UAV supported by a RIS to prepare a platform for  terahertz communications. To this end, UAV’s trajectory, RIS phase-shift, resource allocation, and power control are jointly optimized. However, since the environment contains no uncertainties, the problem is solved using iterative algorithm, namely successive convex approximation. 

The works presented above consider various applications and scenarios for UAVs and RIS deployment. However, to the best of our knowledge, no work in the literature accounted for UAV trajectory planning in 2D space with the support of a discrete RIS to collect data generated by IoTDs with a target deadline.

\section{System Model}
\label{sec-system-model}
We consider IoTDs scattered in an urban environment to collect data essential for one or more of emerging smart cities services. The device alternates between an active and passive activation mode to conserve energy and has a period during which its information should be collected before it becomes stale and bear no value \cite{samir2020trajectory}. Each device location is denoted by ($x_i, y_i, z_i$). A UAV is dispatched to gather data from these devices and has a fixed altitude $z_U$. The UAV can move in two dimensions $x_U^n, y_U^n$. In order to enhance the communication between the IoTDs and UAV, a RIS equipped with $M$ reflecting elements is placed at ($x_R, y_R, z_R$) as demonstrated in Fig. \ref{fig:system_model_iot}. The list of parameters and variables used is shown in Table \ref{table:mathematical_notations}. In this section, two objectives are developed. The first one accounts for maximizing the number of IoT devices served. The second objective optimizes the UAV energy efficiency.

\begin{table}[t]
	\caption{Mathematical notations}
	\begin{center}
		\begin{tabular}{|c|p{6cm}|}
		    \hline
    		\rowcolor{lightgray} \multicolumn{2}{|c|}{Parameters} \\
			\hline
			$I$ & A set of IoTDs existing in the area. \\
			\hline
			$N$ & Time horizon. \\
			\hline
			$(x_U^n, y_U^n, z_U)$ & UAV coordinates at time slots $n$. \\
			\hline
			$(x_R, y_R, z_R)$ & RIS coordinates. \\
			\hline
			$(x_i, y_i, z_i)$ & IoTD $i$ coordinates. \\
			\hline
			$Z_i$& Size of data transmitted by device $i$. \\
			\hline
			$\sigma^2$& thermal noise power. \\
			\hline
			$\rho$ & The mean path gain at reference distance = 1m. \\
			\hline
			$\alpha$& Path loss exponent. \\
			\hline
			$P$& Transmission power. \\
			\hline
			$K$& Rician factor. \\
			\hline
			$\delta_i^n$& Decision variable for wireless scheduling. \\
			\hline
			$T_i$& Active period starting time of IoTD $i$. \\
			\hline
			$F_i$& Active period ending time (deadline) of IoTD $i$. \\
			\hline
			$C$& Number of UAV wireless channels \\
			\hline
			$b$ & Number of control bits for the RIS elements.\\
			\hline
			$Q$ & Number of RIS phase-shift patterns \\
			\hline
			$\phi_{R,U}^n$& Angle of arrival at RIS from UAV at time slot $n$. \\
			\hline
			$\phi_{R,i}$& Angle of arrival between the RIS and IoTD $i$. \\
			\hline
			$\Omega$& Set of available phase shift values for the RIS elements.  \\
			\hline
			$\zeta$& Separation value between RIS elements.  \\
			\hline
			$\lambda$& Carrier wave length.  \\
			\hline
			$\omega_i$& Service indicator equals to 1 is IoTD $i$ served and 0 otherwise.  \\
			\hline
			$\Gamma_i^n$ & signal-to-noise-ration for IoTD $i$ at time slot $n$. \\
			\hline
		\end{tabular}
		\label{table:mathematical_notations}
	\end{center}
\end{table}

\subsection{IoT Activation}
\label{sec-iot-activation}
IoTDs are battery-powered, hence, they are energy constrained \cite{kouzayha2017joint}. Consequently, IoTDs tend to be power-conservative by adopting two operational modes; sleep and active \cite{al2019energy}. The duration and frequency of sleep and active modes depend on the applications, for instance, measuring soil humidity occurs once every hours \cite{kim2020deep} while transmitting autonomous vehicles information happens in a much faster pace to keep the traffic smooth and safe. Other applications may require more frequent patterns of activation. In addition, according to \cite{wang2018access, kozlowski2019energy}, the randomness of IoT device activation pattern can be modeled using a uniform distribution. 

\subsection{Communication Model}
In this section, we present the channel model, the signal-to-noise-ratio (SNR), and the achievable data rate analysis. In our model, we assume that the IoTDs transmit their data to the UAV in the uplink using frequency division multiple access
(FDMA). We denote the total number of available radio resources as $C$.  In addition, the channel model between the IoTDs and the UAV which is referred to as "direct link" and cascaded channel model IoTDs-RIS-UAV which is referred to as "indirect link".

\subsubsection{Direct Link}
We assume a channel model for the UAV in urban area where high-rise buildings and other objects appear which could disturb the links between the UAV and IoTDs. Thus, we assume that the link propagation is characterized by both strong Line-of-Sight (LoS) and non Line-of-Sight (NLoS). Here, $S_{U\to i}^n \in \{$LoS, NLoS$\}$ indicates the state of the channel between the UAV and IoTD $i$ at time slot $n$. The probability of having LoS link adopted in this paper is similar to \cite{mozaffari2016unmanned}.
Then, we can find the probability of channel states between the UAV and IoTD $i$.
\begin{equation}
\begin{split}
    Pr(S_{U\to i}^n=\text{LoS}) = \dfrac{1}{1+\eta_1 e^ {(-\eta_2 (\theta_{U \to i}^n - \eta_1)) }}, \\ \forall i \in I^n, n,
\end{split}
\end{equation}
where $\eta_1$ and $\eta_2$ are constant parameters of the environment. $\theta_{U\to i}^n = \dfrac{180}{\pi} \arctan(\dfrac{z_U}{\widehat{D}_{U\to i}^n})$ is the angle degree between IoTD $i$ and the UAV at time slot $n$. Meanwhile, $z_U$ denotes the height of the UAV antenna and $\widehat{D}_{U\to i}^n$ is the horizontal distance between IoTD $i$ and the UAV at time slot $n$. Moreover, the probability of having NLoS can be obtained from  $Pr(S_{U\to i}^n=\text{NLoS}) = 1 - Pr(\text{$S_{U\to i}^n=$LoS})$. Next, the channel gain for each IoTD $i$ at time slot $n$ is computed as:
\begin{equation}
\boldsymbol{h}_{U, i}^n =\begin{cases}
(D_{U, i}^n)^{-\beta_{1}} & \parbox{10em}{$S_{U\to i}^n=$ LoS,}\\
\beta_2 (D_{U, i}^n)^{-\beta_1} & \parbox{10em}{otherwise,}
\end{cases}
\end{equation}
where $D_{U, i}^n$ is the euclidean distance between the UAV and IoTD $i$ at time slot $n$; $D_{U, i}^n=\sqrt{(\widehat{D}_{U\to i}^n)^2 + z_U^2}$. $\beta_1$ denotes the path loss exponent and $\beta_2$ is the attenuation factor for NLoS. Thus $\boldsymbol{h}_{U,i}^n$ can also be rewritten as:
\begin{multline}
    \boldsymbol{h}_{U,i}^n= Pr(S_{U\to i}^n=\text{LoS}) (D_{U, i}^n)^{-\beta_1} + \\ (1 - Pr(S_{U\to i}^n=\text{LoS})) \beta_2 (D_{U, i}^n)^{-\beta_1}    
\end{multline}

\subsubsection{Indirect Link}
We consider a uniform linear array (ULA) RIS \cite{long2020reflections}. In addition, similar to the UAV, the RIS is assumed to have a certain height, $z_I$.  The communication links between the UAV and RIS and that between the RIS and IoTD $i$ are assumed to have a dominant line-of-sight (LoS). Thus, these communication links experience small-scale fading which are modeled as Rician fading with pure LoS components \cite{abdullah2020hybrid, samir2020optimizing}. Consequently, the channel gain between the UAV and RIS, $\boldsymbol{h}_{R,U} \in \mathbb{C}^{M \times 1}$, can be formulated as follows.

\begin{equation}
    \boldsymbol{h}_{R,U}^n = \underbrace{\sqrt{\rho (D_{R, U}^n)^{-\alpha}}}_{\mathrm{path~loss}}  \underbrace{\sqrt{\dfrac{K}{1+K}} \boldsymbol{\bar{h}}_{R, U}^{n, \mathrm{LoS}}}_{\mathrm{Rician~fading}},
\end{equation}
where $D_{R, U}^n$ is the Euclidean distance between the RIS and UAV and can be computed from $\sqrt{(x_R - x_U^n)^2 + (y_R - y_U^n)^2 + (z_R - z_U)^2}$. Also, $\rho$ is the average path loss power gain at reference distance $D_0=1$m. Also, $K$ is the Rician factor and $\boldsymbol{\bar{h}}_{R, U}^{\mathrm{LoS}, n}$ is the deterministic LoS component which can be defined as follows 
\begin{equation}
\begin{split}
\label{eq:channel-gain-I-R}
\boldsymbol{\bar{h}}_{R, U}^{\mathrm{LoS}, n} = \underbrace{\Bigg[1, e^{-j\frac{2\pi}{\lambda} \zeta \phi^n_{R, U}},..., e^{-j\frac{2\pi}{\lambda} (M-1) \zeta \phi^n_{R, U}}\Bigg]^{T}}_{\mathrm{array~response}},\\ \forall n \in N,
\end{split}
\end{equation}
where $\phi_{R, U}^n=\dfrac{x_R - x_U^n}{D_{R, U}^n}$ is cosine of the angle of arrival of signal from the RIS to UAV. $\zeta$ is the separation between RIS elements and $\lambda$ is the carrier wavelength.  

Similarly, we can compute the channel gain between the RIS and IoTDs which is denoted by $\boldsymbol{h}_{R,i}^n \in \mathbb{C}^{M \times 1}$. 
\begin{equation}
\label{eq:channel-gain-I-v}
    \boldsymbol{h}_{R, i}^n = \underbrace{\sqrt{\rho (D_{R, i})^{-\alpha}}}_{\mathrm{path~loss}}  \underbrace{\sqrt{\dfrac{K}{1+K}} \bar{h}_{R, i}^{n~\mathrm{LoS}}}_{\mathrm{Rician~fading}}, \forall i, n \in N, 
\end{equation}

\begin{equation}
\begin{split}
  \boldsymbol{\bar{h}}_{R, i}^{n~\mathrm{LoS}} = \underbrace{\Bigg[1, e^{-j\frac{2\pi}{\lambda} \zeta \phi_{R, i}},..., e^{-j\frac{2\pi}{\lambda} (M-1) \zeta \phi_{R, i}}\Bigg]^{T}}_{\mathrm{array~response}}, \\ \forall i, n \in N,  
\end{split}
\end{equation}
where $D_{R, i}$ is the euclidean distance between the RIS and IoTD $i$ and $\phi_{R, i}=\dfrac{x_R - x_i}{D_{R, i}}$.

Denote the phase-shift matrix of the RIS in the $n$th time slot as $\boldsymbol{\Theta}^n = \diag\{e^{j\theta_1^n},..., e^{j\theta_M^n}\}$, where $\theta_m^n$ is the phase-shift of the $m$th reflecting element $m = 1, 2, · · · , M$. Due to the hardware limitations, the phase-shift can only be selected from a finite set of discrete values. Specifically, the set of discrete values for each RIS reflecting element can be given as $\theta_m^n \in \Omega = \{0, \frac{2\pi}{Q}, \dots, \frac{2\pi(Q - 1)}{Q}\}$, where $Q = 2^b$ and $b$ is the number of bits that control the number of available phase-shifts for the RIS elements. Hence, the SNR is:
\begin{equation}
\label{eq:snr}
\Gamma^n_i = \dfrac{P \abs{\boldsymbol{h}_{U,i}^n + \boldsymbol{h}_{R, U}^{n,H}  \boldsymbol{\Theta}^n \boldsymbol{h}_{R, i}^n}^2}{\sigma^2}, \forall i, n \in N,
\end{equation}
\newcounter{mytempeqncnt}
where $P$ is the IoTD transmit power and $H$ denotes the conjugate transpose operator. Next, we can compute the channel gain component as in Eq \eqref{eq:snr-complete}.
\setcounter{equation}{9}

\begin{figure*}[!t]
\normalsize
\setcounter{mytempeqncnt}{\value{equation}}
\setcounter{equation}{8}
\begin{equation}
\label{eq:snr-complete}
\begin{split}
\boldsymbol{h}_{U\to i}^n + \boldsymbol{h}_{R, U}^{n,H}  \boldsymbol{\Theta}^n \boldsymbol{h}_{R, i}^n = Pr(S_{U\to i}^n=\text{LoS}) (D_{U\to i}^n)^{-\beta_1} + \beta_2 (D_{U\to i}^n)^{-\beta_1} -  Pr(S_{U\to i}^n=\text{LoS}) \beta_2 (D_{U\to i}^n)^{-\beta_1} \\ + \dfrac{\rho \dfrac{K}{K+1}}{\sqrt{(d_{R,i})^{\alpha}} \sqrt{(d_{R,U}^n)^{\alpha}}}  \times \sum_{m=1}^M e^{j(\theta_m^n + \frac{2\pi}{\lambda} (m-1) \zeta \phi_{R,i} - \frac{2\pi}{\lambda} (m-1) \zeta \phi^n_{R,U})}, \forall i, n \in N.
\end{split}
\end{equation}
\setcounter{equation}{\value{mytempeqncnt}}
\hrulefill
\vspace*{4pt}
\end{figure*}

The amount of data collected from each IoTD at time slot $n$ can then be computed as follows.
\begin{equation}
l_i^n = \delta_i^n \log_2(1 + \Gamma_i^n), \forall i, n,
\end{equation}
where $\delta_i^n$ is a scheduling decision variable (1 if device $i$ is scheduled at time slot $n$ and 0 otherwise).

\subsection{UAV Service Rate}
In this section, we seek to maximize the number of served IoTs, each during its activation period subject to several operational constraints.
\allowdisplaybreaks
\begingroup
\begin{maxi!}|s|[2]
    {\boldsymbol{\Theta}, x_U, y_U, \delta}
    { \sum_{i=1}^{I} \omega_i \label{eq:objective}} 
    {\label{eq:Example1}} 
    {\text{($\mathcal{P}$1)}\quad}
    \addConstraint{}{\sum_{n=1}^{N}l_i^n \geq \omega_i Z_i \label{eq:con1}}
    \addConstraint{}{\theta_m^n \in \Omega, \forall n \in N, m \in M \label{eq:con2}}  
    \addConstraint{}{\sqrt{(x_U^{n+1} - x_U^n)^2 + (y_U^{n+1} - y_U^n)^2} \leq V
    \label{eq:con3}}
    \addConstraint{}{0 \leq x_U^n \leq X
    \label{eq:con4}}
    \addConstraint{}{0 \leq y_U^n \leq Y
    \label{eq:con5}}
    \addConstraint{}{\sum_{i=1}^{I} \delta_i^n \leq C, \forall n \in N. \label{eq:con6}}  
    \addConstraint{}{\delta_i^n \leq 1 + \dfrac{n - T_i - 1}{O}, \forall i \in I \label{eq:con7}}
    \addConstraint{}{\delta_i^n \leq 1 + \dfrac{F_i - n}{O} , \forall i \in I, \label{eq:con8}}
\end{maxi!}
\endgroup
where $O$ is a large number. In Eq \eqref{eq:con1}, $\omega_i$ indicates whether a IoTD data has been collected successfully ($\omega_i=1$) or not ($\omega_i=0$), therefore, the objective aims at maximizing the number of IoTDs admitted for service. Eq \eqref{eq:con2} ensures that phase shifts are set within their possible range. Eq \eqref{eq:con3} does not allow the UAV to violate its maximum speed. Eq \eqref{eq:con4} and \eqref{eq:con5} refrain the UAV from leaving the concerned area. Eq \eqref{eq:con6} emphasizes that the number of IoTD scheduled to transmit at each time slot does not violate the available resources. Eq \eqref{eq:con7} and \eqref{eq:con8} make sure that the IoTDs are only scheduled to transmit their data during the active period.

The above problem is a mixed-integer non-convex owing to the existence of integer and binary decision variables of wireless scheduling and RIS discrete phase shift. Besides, the UAV trajectory is non-convex \cite{samir2019uav}. Hence, the resulted problem is difficult to solve or obtain its optimal solution in polynomial time. Further, some environment parameters such as IoTDs' active periods are not given in real-world. In this sense, dynamic programming and linear programming based solutions are not suitable to solve such problems. Therefore, we resort to DRL to tackle the problem where DRL has been used widely to effectively solve similar problems. First, we cast the original problem into two sub-problems. The first part associates with UAV trajectory and IoTDs scheduling. This part is formulated as MDP and then solved via PPO agent. The output of the first part is then fed to the second sub-problem to solve the RIS phase-shift configuration via iterative BCD. The complete solution will be discussed next.

\subsection{UAV Energy Efficiency}
One of the main resource limitation of UAVs is the battery capacity. Energy consumption of the UAV results from two sides; namely, UAV mobility and wireless communication. The first part, associated with UAV mobility, constitutes the dominant share of energy consumed from the UAV battery. Subsequently, in this section, we focus only on this part.

For this study, we use the metric of energy efficiency to quantify the benefits of RIS in aiding the UAV. In short, energy efficiency advocates for smaller energy consumption in regards to perform higher. This can save device's battery life and improve energy consumption \cite{dechene2012energy}. In the context of this article, this translates to uploading more data while consuming less energy and can be expressed as:

\allowdisplaybreaks
\begingroup
\begin{maxi!}|s|[2]
    {\boldsymbol{\Theta}, x_U, y_U, \delta}
    { \dfrac{\sum_{i=1}^{I} s_i Z_i}{E(\nu)_{total}} \label{eq:objective2}} 
    {\label{eq:Example2}} 
    {\text{($\mathcal{P}$2)}\quad}
    \addConstraint{}{\text{Constraints}~ \eqref{eq:con1} - \eqref{eq:con8}}
\end{maxi!}
\endgroup
where $E(\nu)_{total}$ represents the total energy consumption of the UAV throughout its operational time. The mobility of the UAV incurs energy and the amount consumed depends on the velocity of the UAV \cite{zeng2019energy, samir2020trajectory} as in Eq(\ref{eq:uav_power_consumption}).

\newcounter{mytempeqncnt2}
\setcounter{equation}{13}

\begin{figure*}[!t]
\normalsize
\setcounter{mytempeqncnt2}{\value{equation}}
\setcounter{equation}{12}
\begin{equation}
\label{eq:uav_power_consumption}
P(\nu^n)_{total} =  \underbrace{G\bigg(1+3\dfrac{M_{tip}^2}{w_b^2}\bigg)}_\text{Blade profile power}  +  \underbrace{\dfrac{1}{2} \pi (\nu^n)^3 H}_\text{Parasite power} \\ +  \underbrace{m_u g \sqrt{\Bigg( \dfrac{-(\nu^n)^2 + \sqrt{(\nu^n)^4 + ( \dfrac{m_U g}{\pi A} )^2}}{2} \Bigg)}}_\text{Induced power}.
\end{equation}
\setcounter{equation}{\value{mytempeqncnt2}}
\hrulefill
\vspace*{4pt}
\end{figure*}
Where $\nu^n$ denotes UAV velocity at time slot $n$ and $M_{tip}$ is the blade’s rotor speed, $G$ and $H$ are two constants which depend on the dimensions of the blade and the UAV drag coefficient, respectively, $\pi$ is the air density, $m_U$ and $g$ respectively denote the mass of the UAV and the standard gravity, $A$ is the area of the UAV. The total energy consumption to cover a distance $d$ at a constant velocity UAV $w$ can be computed as $E(\nu)_{total} = \int^{d/\nu}_{0} P(\nu) dt = P(\nu)\dfrac{d}{\nu}$ as in \cite{samir2020trajectory}.

\textbf{Remark:} This article considers the UAV as a mobile data collector to gather/process data generated by IoT devices in a particular area and within a comparatively short period of time. It is worth noting that there are significant investments in UAVs to improve the UAVs' battery performance to allow them to fly for longer periods. For example, "Impossible US-1", a UAV model, flies up to 70 minutes \cite{impossible}. Moreover, "Scout B-330" can fly for around 180 minutes \cite{fotouhi2019survey}. In addition, there exists several techniques that can be leveraged to maximize the flight endurance of the UAV \cite{fotouhi2019survey}. For example, solar panels can be mounted on the UAV and the battery is used as a backup when the solar cells are unable to produce enough power while flying under clouds or any zone where the sunlight is blocked. Generally, there are several technologies that can provide drones with very long flight endurance to handle the energy limitations of most UAVs. Therefore, this work concentrates primarily on maximizing the service offered to IoT devices within a certain time horizon by optimizing its trajectory, wireless scheduling and RIS configuration. Therefore, $\text{($\mathcal{P}$1)}$ will be our main objective to solve while the impacts of RIS on the UAV energy efficiency, $\text{($\mathcal{P}$2)}$, will also be examined in the numerical results.

\section{Solution Approach}
\label{sec-solution-approach}
\subsection{UAV Mobility}
First, the problem defined in the previous section is converted to MDP by defining the 5-tuple $\langle\boldsymbol{S}, \boldsymbol{A}, \gamma , \boldsymbol{R}, \boldsymbol{P}\rangle$ where:
    \begin{itemize}

    \item $\boldsymbol{S}$ is a set of states, also known as state space, that includes all the possible states $s^n \in \boldsymbol{S}$ at any time slot $n$. 
    
    \item $\boldsymbol{A}$ is a set of possible actions, also known as action space, that an agent can take at each time slot $n$ which is denoted by $a^n$.
    
    \item $\gamma$ is the discount factor satisfying $0 \leq \gamma < 1$ and it specifies how much the decision maker cares about rewards in the distant future relative to those in the immediate future. 
    
    \item $\boldsymbol{P}$ is the transition probability of being in next state given the current state and current action $Pr(s^{n+1} \mid s^n , a^n), \forall s^{n+1}, s^{n} \in \boldsymbol{S}, a^{n} \in \boldsymbol{A}$.
    
    \item $\boldsymbol{R} : \boldsymbol{S} \times \boldsymbol{A} \rightarrow \mathbb{R}$ is a reward function where $r^n = r(s^n, a^n, s^{n+1})$ denotes the single-step reward of the system for transitioning from state $s^n$ to state $s^{n+1}$ due to action $a^n$.
    \end{itemize}

where the state, action, reward, are defined as follows:
\begin{itemize}
    \item State $\boldsymbol{S}$: The state at time slot $n$, $s^n \in \boldsymbol{S}$, can be expressed by:
    
    \begin{equation}
        \label{eq:state}
        s^n = [x_U^n, y_U^n, \boldsymbol{X}, \boldsymbol{Y}, \boldsymbol{T}^n, \boldsymbol{F}^n, \boldsymbol{U}^n, \boldsymbol{Z}^n].
    \end{equation}
    
    Where the state $s^n \in S$  includes UAV position at time slot $n$ ($x_U^n, y_U^n$). Two variables $\boldsymbol{X}, \boldsymbol{Y}$ denote the IoTDs' locations. A vector $\boldsymbol{T}^n$ represents active period starting time for all the IoTDs at time slot $n$. A vector $\boldsymbol{F}^n$ represents active period deadline for all the IoTDs at time slot $n$. A vector $\boldsymbol{U}^n$ represents the total amounts of data fetched from the IoTDs by time slot $n$. A vector $\boldsymbol{Z}^n$ represents the amount of data to be uploaded from all the IoTDs at time slot $n$.
    
    \item Action $\boldsymbol{A}$: The action taken at time slot $n$, $a^n \in \boldsymbol{A}$ contains two sub-actions. First, $a_U^n$, is the UAV trajectory ($x_U, y_U$). Second, $a_i^n$, is the IoTD wireless scheduling. In order to handle the first part of the action space which is related to UAV mobility that is continuous for both distance and direction of movement (angle) in 2-D space, we discretize this part into 5 directions related to the different actions that represent all kinds of movements that the UAV can take (left, right, forward, backward, stop). The second part of the action space determines the IoTDs wireless scheduling. To do so, the agent has to determine which IoTDs will be scheduled to transmit in the current time slot, meaning, the agent has to select a subset of size $C$ from $I$. However, not all the IoTDs can transmit at each time slot since some of them are in sleep mode or have already uploaded their data. The average number of IoTD that are active in one time slot can be determined from the activation pattern of the IoTDs, which is chosen to be uniform in this work, and can be expressed as $I\Lambda/N$ where $\Lambda$ represents the average active period length of the IoTDs\footnote{For the sake of simplicity, in this work we assume all IoTDs have the same active period length.}. Eventually, the total number of actions available can be realised through:
    
    \begin{equation}
         5 \times \dfrac{\Big(\dfrac{I\Lambda}{N}\Big)!}{C! \Big(\dfrac{I\Lambda}{N}-C\Big)!},
    \end{equation}
    where 5 corresponds to the UAV trajectory and the second part represents all the combinations of IoTDs that can be scheduled to serve in one time slot.
    
    \item Reward: The immediate reward, $r^n$, is equal to the number of IoTDs served if the respective data has been completely uploaded to the UAV at time slot $n$. Otherwise, $r^n$ is set to 0.
    
\end{itemize}

For DRL, we exploit PPO to develop our agent as laid out in Algorithm \ref{algorithm:ppo_algorithm}. First, the agent initializes random sampling policy and value function for the neural networks as in lines 3 and 4. Then, in each epoch, the agent observes the environment which consists of the set of IoTDs and their information, active periods, data uploaded, etc. Then at each time slot $n$, the agent decides where the UAV should go next. If the next movement is still within the area of interest, the UAV will move, otherwise, the UAV stays at its current location. This will make sure that the UAV does not leave the area and will help the agent to learn faster by having less states and actions to explore. Next, the agent selects an action which is a binary vector that determines which set of IoTDs will be served via the UAV. Based on that action, the BCD algorithm is then invoked to configure the phase-shift matrix in order to maximize the channel gain. Eventually, the instantaneous reward is worked out which has two cases. First, if no IoTD has uploaded its value, the step reward will be set to 0. Otherwise, the step reward will be equal to the number of IoTDs that has managed to upload their data to the UAV completely at that time slot.

After gathering the set of samples and computing the rewards, the agent works out the advantage function (line 15), $\hat{A}$, which is defined as the resultant of subtracting the expected value function from the actual reward. $\hat{A}$ is the estimated advantage function or relative value of the selected action. It helps the system to understand how good it is preforming based on its normal estimate function value \cite{bohn2019deep}.
    \begin{center}
    \begin{algorithm}[t]
    	\caption{Proposed DRL for Scheduling}
    	\label{algorithm:ppo_algorithm}
    	{
    	    \textbf{Inputs:} $N$, $v$, Learning Rate, $\gamma$, $\epsilon$.\\
    	    \textbf{Outputs:} UAV trajectory.\\
    	    Initialize policy $\pi$ with random parameter $\theta$ and threshold $\epsilon$\\
    	    Initialize value function $V$ with random parameters $\phi$\\
    	    \For{each episode $k \in \{0, 1, 2,...\}$}
    	    {
    	        \While{$n < N$}
    	        {
    	            Set $n = n+1$.\\
    	            Observe state $s^n$.\\
    	            Choose action $a^n$. \\
    	            Move the UAV based on $a_U^n$. \\
    	            \If{UAV new location is not  outside the area of interest}
    	            {
    	                Keep the UAV at its last location.\\
    	            }
    	            Schedule the IoT devices based on $a_i^n$.\\
    	            Compute the reward $r^n$.
    	        }
    	        Compute advantage estimate $\hat{A}$ for all epochs.\\
    	        Optimize surrogate loss function using Adam optimizer.\\
    	        Update policy $\pi_{\theta_{old}} \gets \pi_{\theta}$.\\
    	    }
    	}
    \end{algorithm}
    \end{center}
\vspace{-0.9cm}

\subsection{A BCD Method for RIS Configuration}
After the DRL agent has decided the next UAV move as well as the set of scheduled IoTDs at time slot $n$, Block Coordinate Descent (BCD) is then invoked to tune the RIS phase-shift coefficients. The objective of the BCD is to maximize the amount of data collected at each time slot by maximizing the achievable data rate through all the scheduled IoTDs which can be expressed as:

\begingroup
\small
\begin{maxi!}|s|[2]
    {\boldsymbol{\Theta}}
    {\sum_{i=1}^{I} \delta_i^n \log_2(1 + \dfrac{P \abs{\boldsymbol{h}_{U\to i}^n + \boldsymbol{h}_{R, U}^{n,H}  \boldsymbol{\Theta}^n \boldsymbol{h}_{R, i}^n}^2}{\sigma^2}), n. \label{eq:max-sum-bcd}} 
    {\label{eq:Example111}} 
    {} 
    \addConstraint{}{\theta_m^n \in \Omega, \forall n \in N, m \in M. \label{eq:con2222}}  
\end{maxi!}
\endgroup

\begin{figure*}[htbp]
	\centerline{\includegraphics[scale=0.40]{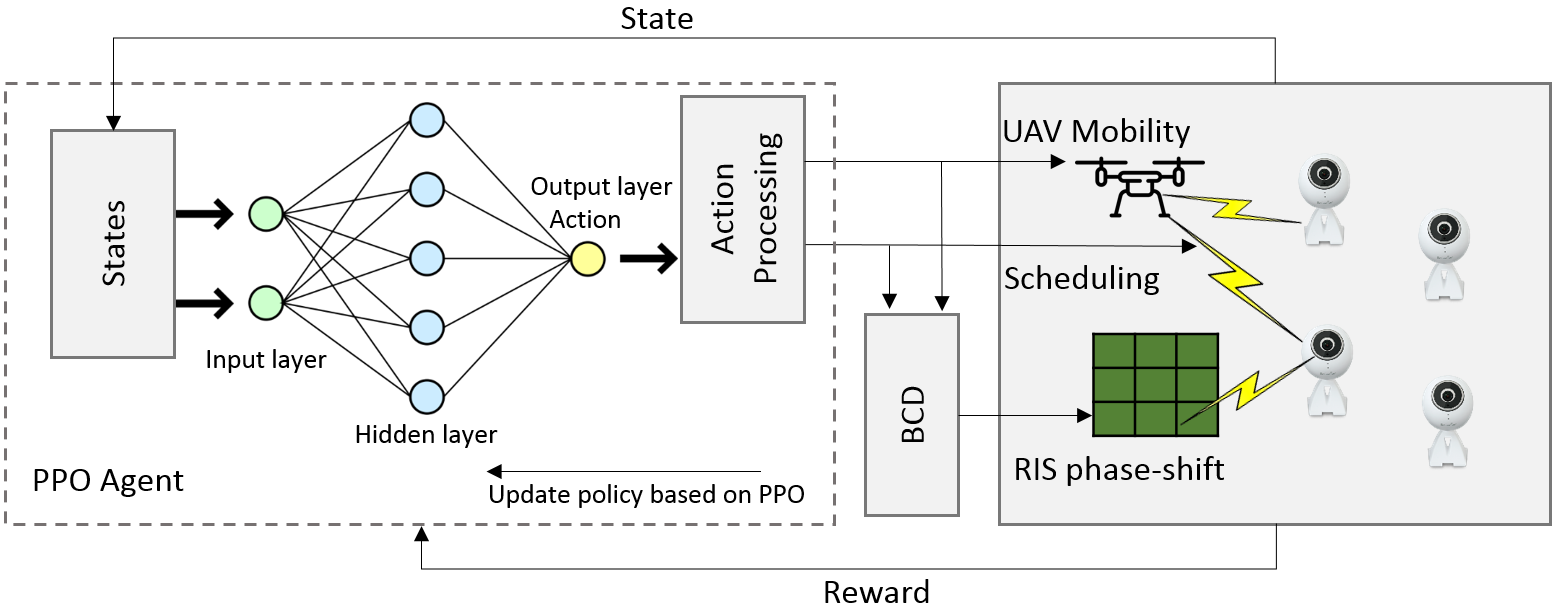}}
	\caption{Solution architecture.}
	\label{fig:solution-architecture}
\end{figure*}

The BCD algorithm works in an iterative way where a sequence of block optimization procedures are performed. That is, in each iteration, it optimizes one RIS element by checking all the possible values in $\Omega$ while fixing the other elements. The value that maximizes the objective in Eq \eqref{eq:max-sum-bcd} will be chosen. After that, the next element will be selected to be optimized and so forth until all the RIS elements are optimized. 
This procedure repeats until there is no change in the RIS phase-shift configuration. The main details of the proposed BCD approach are presented in Algorithm \ref{algortihm:BCD}. Regarding the complexity of Algorithm \ref{algortihm:BCD}, it is $O(V M 2^b)$ where $V$ represents the number of iterations needed until Eq \eqref{eq:max-sum-bcd} is maximized. Indeed, there are three loops embedded in Algorithm \ref{algortihm:BCD}; the first loop corresponds to the number of required iterations, the second one is to iterate over each RIS element, and the last one denotes the number of quantization levels to control each element. For the sake of illustration, the complete solution approach is sketched in Fig. \ref{fig:solution-architecture} where the two main components, PPO and BCD, are shown. After the PPO agent observes the state of the environment, it selects an action with the help of neural networks. The action is then processed to find an action for the UAV mobility and an action for wireless resource scheduling. The two sub-actions are then fed into the BCD algorithm to tune the RIS phase shift. Eventually, the environment computes the step-reward resulted from that action and sends it back to the agent.
%
    \begin{center}
    \begin{algorithm}[t]
        \small
    	\caption{Configure The RIS Elements}
    	\label{algortihm:BCD}
    	{
    	\textbf{Inputs:} $\delta_i^n$.\\
        \textbf{Outputs:} $\boldsymbol{\theta^n}$\\
        \While{Eq (\ref{eq:max-sum-bcd}) not maximized}
        {
            \For{$m=1,...,M$}
            {
                Fix $\theta_{m'}^n, \forall m' \neq m, m' \in M$\\
                Set $\theta_m^n = \underset{\Omega}{\argmax}~ $ Eq(\ref{eq:max-sum-bcd})\\
            }
            Obtain $\theta^n$\\
        }
    	}
    \end{algorithm}
    \end{center}
\section{Numerical Results}
\label{sec-numerical-results}
In this section, we present the simulation results to shed light on the performance of the RIS-assisted UAV in IoT wireless networks. For the simulation settings, we use PPO to build our agent, which is done using Python and TensorFlow. For the DRL, 3 linear layers are used with \emph{tanh} as activation function for the middle layers and \emph{softmax} for the output layer. Internal layers contain 64 units each and \emph{Adam} optimizer is incorporated to minimize the loss function. The learning rate is set to 0.002, $\gamma$ to 0.08, and clip to 0.02. For consistency and accuracy, the results were averaged over 500 data tests. The IoTDs are generated and placed based on a uniform distribution, and their active periods are uniformly distributed. The remaining parameters used in our study are listed in Table \ref{table:expermint-parameters} (unless otherwise indicated).

\begin{table}[htbp]\small
	\caption{Simulation Parameters}
	\begin{center}
		\begin{tabular}{|c|c|}
		    \hline
		    \rowcolor{lightgray} Parameter& Value \\
		    \hline
		    Area size& 300 $\times$ 300 $m^2$\\
		    \hline
		    Number of IoTDs ($I$)& 50 IoTDs\\
			\hline
		    $\sigma^2$ & $- 110$ dBm \\
			\hline
			$K$& 10 dB \cite{abdullah2020hybrid}  \\
			\hline
			$\alpha$& 4\\
			\hline
			$P$& 20 dBm \\
			\hline
			$C$& 3 \\
			\hline
			$\rho$ & 10 dBm\\
			\hline
			$Z_i$ & 50 bits \\
			\hline
			$M$ & 100 \\
			\hline
			$b$ & 2 bits \\
			\hline
			$z_U$& 50 m \\
			\hline
			$z_R$& 1 m \\
			\hline
			$z_i$& 1 m \\
			\hline
			IoT data lifetime & 10 sec\\
			\hline
			$\eta_1, \eta_2$ (for dense urban area) & 11.95, 0.136\\
			\hline
			\end{tabular}
		\label{table:expermint-parameters}
	\end{center}
    \end{table}

\par
\begin{figure}[htbp]
	\centerline{\includegraphics[scale=0.25]{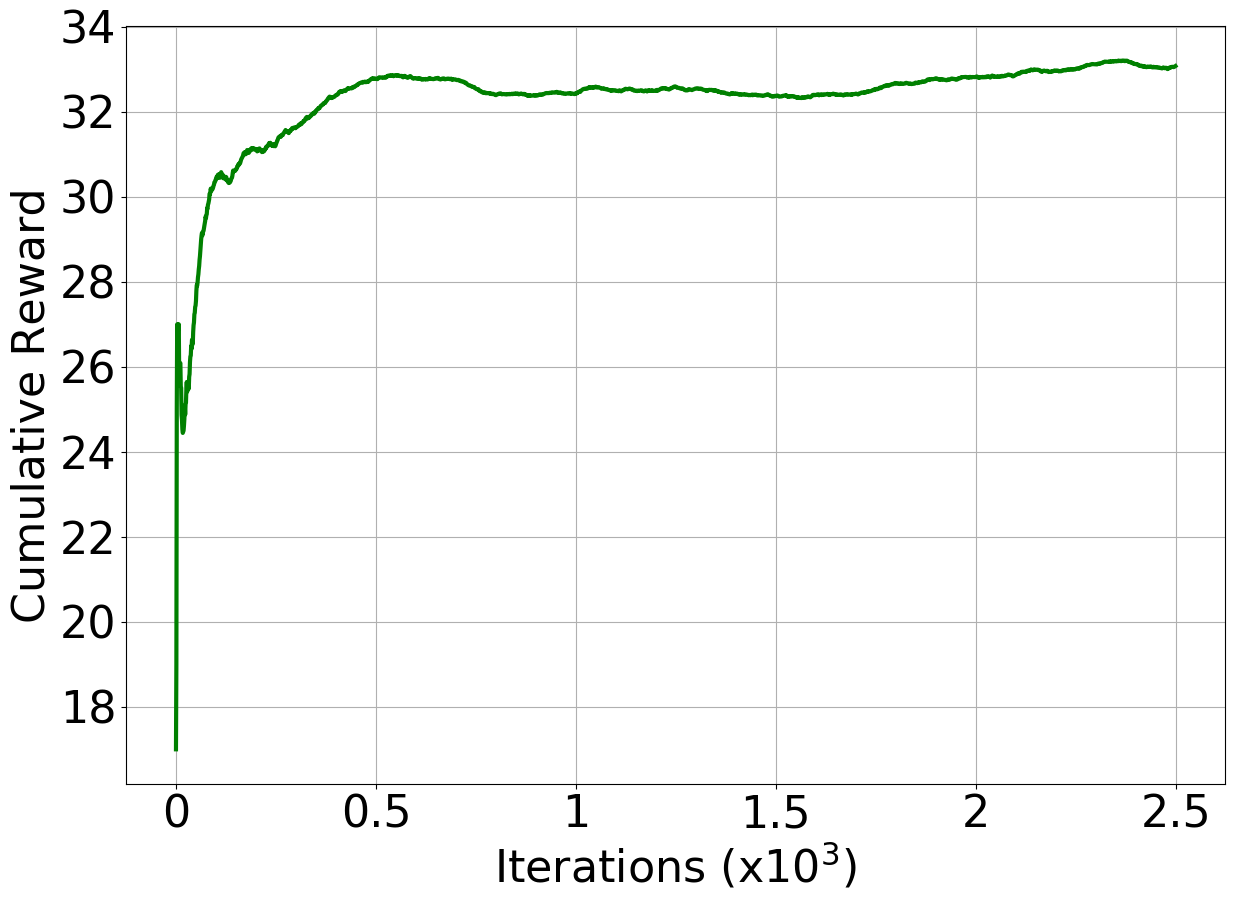}}
	\caption{DRL agent convergence.}
	\label{fig:result_convergence}
\end{figure}

 We observe first the behaviour of the DRL agent; as illustrated in Fig. \ref{fig:result_convergence}, the cumulative reward, which represents the number of IoTDs served, steadily increases as the agent is exposed to more epochs/iterations. We also observe that after around 500 iterations, the system starts to converge.

\begin{figure*}[htbp]
    \centering
    \subfloat[]{\includegraphics[scale=0.27]{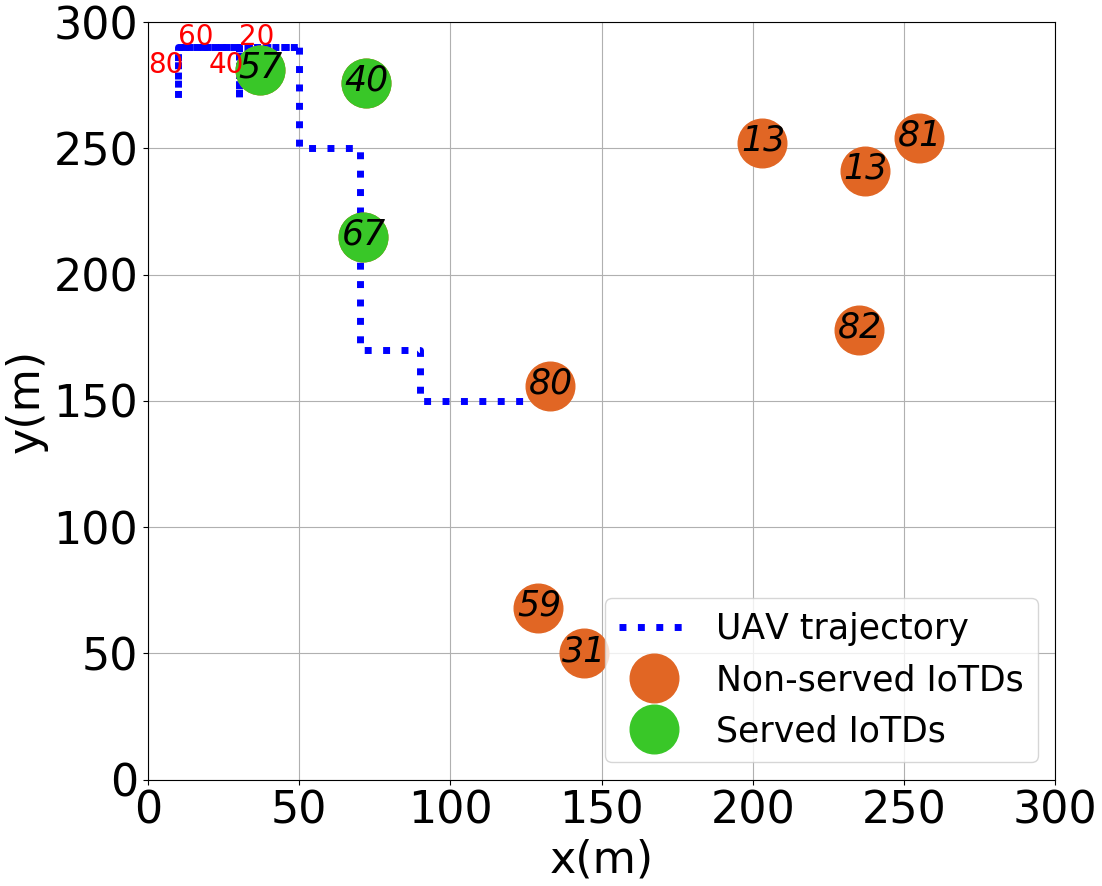}} 
    \subfloat[]{\includegraphics[scale=0.27]{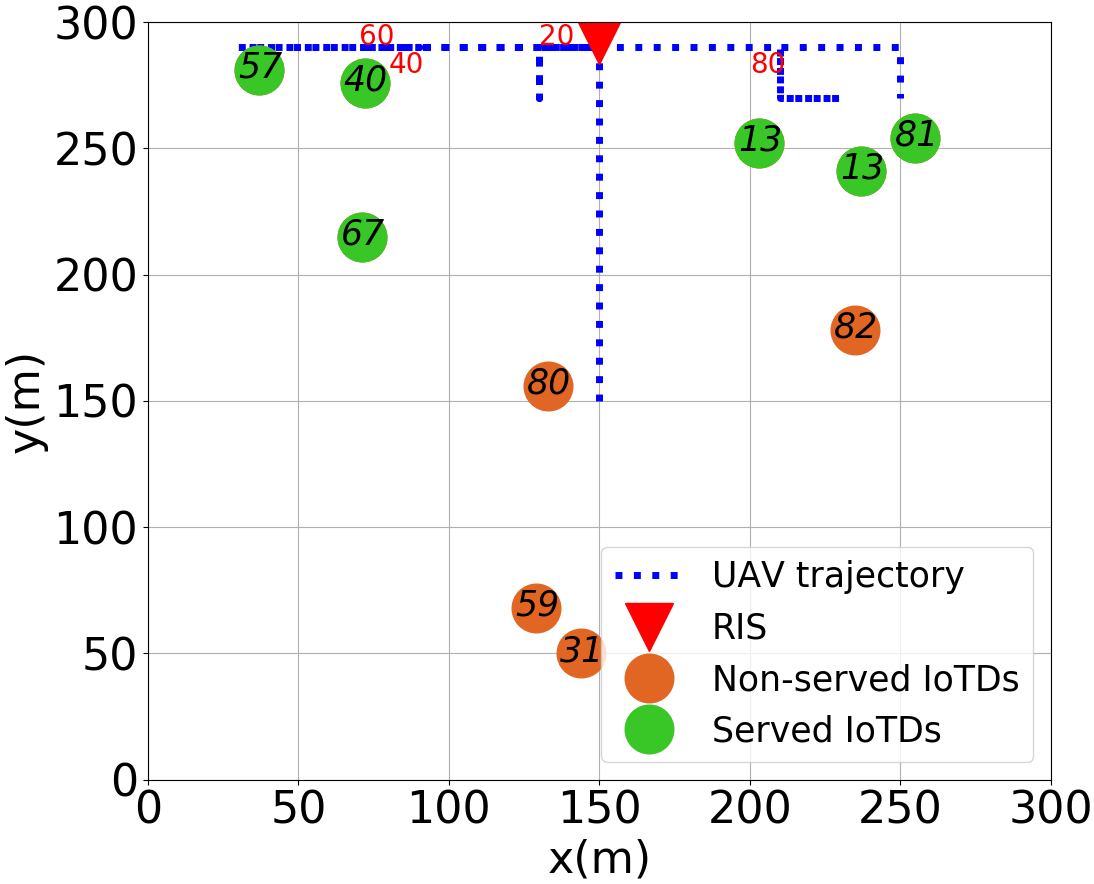}} 
    \caption{UAV trajectory in two scenarios: (a) without RIS (b) with RIS ($M=50$)}
    \label{fig:uav-trajectory}
\end{figure*}

In addition, we show in Fig. \ref{fig:uav-trajectory} the UAV trajectory where two scenarios are considered; without RIS and with RIS ($M=50$). The deadline of the activation periods are written inside the IoTD circle and the numbers in red denote the UAV location at that time slot. 10 IoTDs are considered in this network where each one generates data of size 60 bits. In both cases, DRL is used to control the UAV trajectory. The difference between the two scenarios in terms of mobility behaviour is apparent. In the scenario without RIS, Fig. \ref{fig:uav-trajectory} (a), the UAV flies towards a cluster of IoTDs that have their deadline in sequence and not far from each other. Recall that the UAV has limited resources and cannot collect all the generated data, hence, it is better to select a subset of IoTDs which can be served to completion. In this instance, IoTDs with deadlines 40, 57, and 67, would be the perfect choice as they locate near to each other and their deadlines do not overlap. The rest of IoTDs, such as the one near the center (deadline=80) and the two on the right side (deadline=81 and 82), will remain unserved as the UAV will not have enough time to fly and collect their data before expiration. On the other hand, in Fig. \ref{fig:uav-trajectory} (b), the UAV is able to serve 6 (twice more) devices with the assistance of the RIS. One can also observe that the behaviour of the agent is completely different. For example, first the UAV flies towards the RIS to improve the quality of the indirect LoS. Throughout this travel, the UAV is able to serve two IoTDs with deadline of 13 located on the right side. Then, it positions itself in the left side while balancing the channel quality of the direct and indirect LoS to serve the IoTDs with deadlines 40, 57, and 67. Finally, the UAV flies back to the other side in order to serve the IoTD of deadline 81. We can see that the IoTDs that are located far from the RIS are not served since they have poor indirect LoS and the UAV has restricted maximum speed.

Next, in order to validate the performance of our solution approach, we develop four baseline methods to compare with. Indeed, to the best of our knowledge, no work in the literature addressed similar problem. These are explained as follows.
\begin{itemize}
    \item Random Walk UAV: In this method, the UAV trajectory is determined based on random movement. On the other hand, the phase shift tuning is selected using the BCD method.
    \item Stationary UAV: the UAV is assumed to stay still in the middle of the network while the RIS phase shift is tuned using BCD.
    \item Random $\Theta^n$: the RIS phase shift is configured following random distribution at each time slot while the UAV trajectory is planned using a DRL agent. 
    \item Without RIS: In this method, the UAV has only direct link to communicate with the IoTDs. Again, in this scenario, the UAV is moved using a DRL agent. 
\end{itemize}

In the rest of this section, we study the impact of various environment settings and parameters on the performance of the system and solution approach.

\subsection{The impact of the number of RIS's elements}
\begin{figure}[htbp]
	\centerline{\includegraphics[scale=0.25]{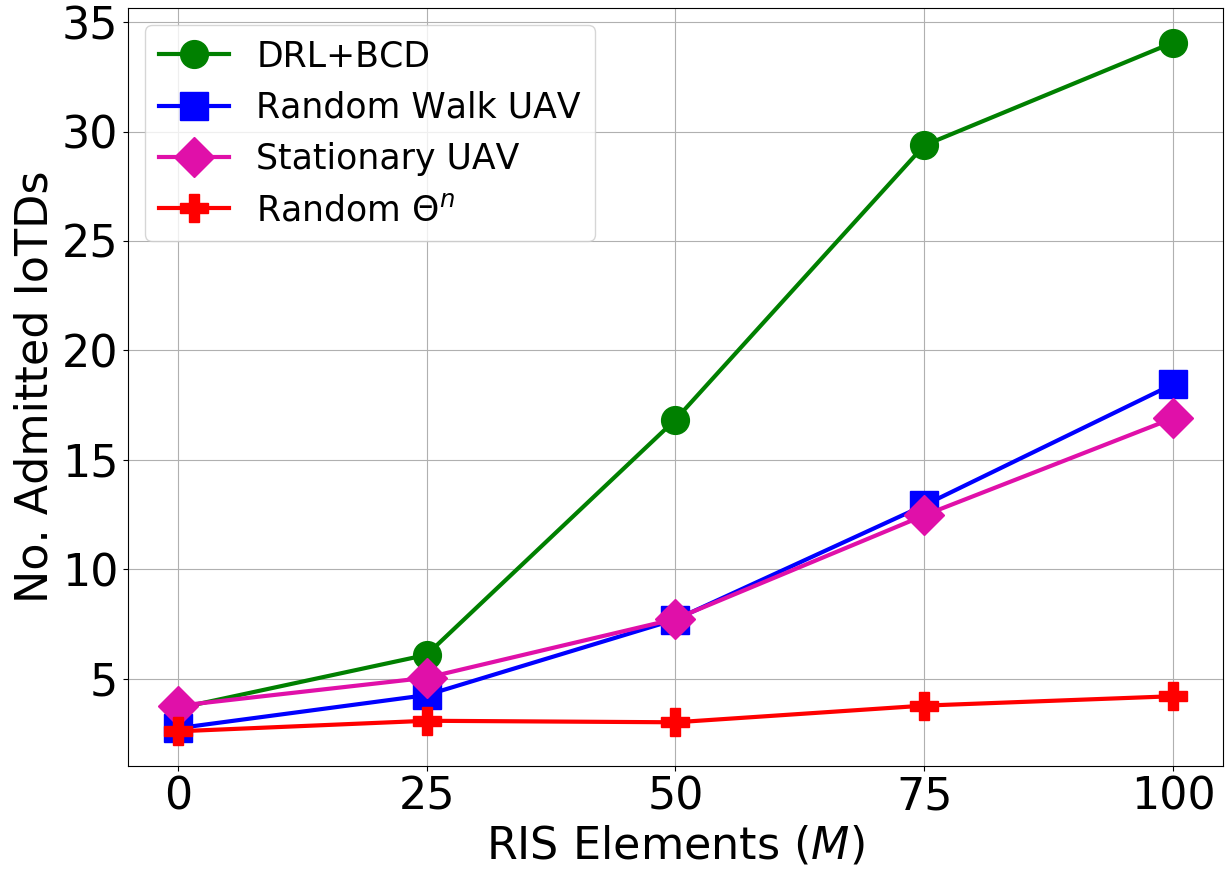}}
	\caption{Effect of the RIS number of elements.}
	\label{fig:result_no_elements}
\end{figure}

Fig. \ref{fig:result_no_elements} shows the number of served devices (data collected within deadline) as we vary the size of the RIS (number of elements). We observe first, that indeed the size of the RIS helps improve the performance and accordingly, more devices are served when more elements are present. Another observation from this figure is that when the phase shifts are randomly selected, not much gain can be seen from the RIS. However, as the RIS elements are optimized, the gain is apparent and varies with the way the UAV trajectory is optimized. Clearly, the best performance is attained when the two are jointly optimized, yielding close to $7\times$ gain when $M=100$. Having a stationary UAV or a UAV with random trajectory have some gains, however the gain is much smaller than when the trajectory is determined by the developed DRL agent.



\subsection{Effect of Network Size}
\begin{figure}[htbp]
	\centerline{\includegraphics[scale=0.25]{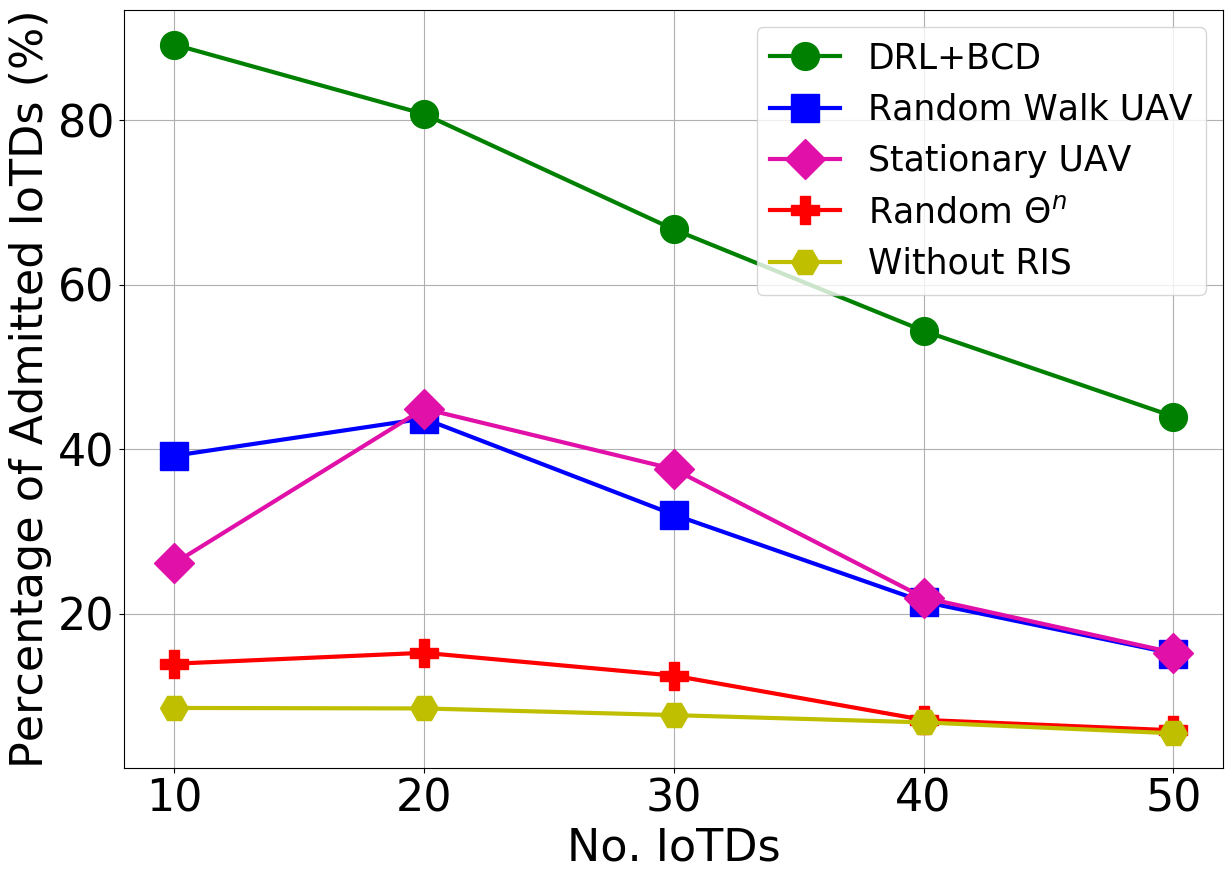}}
	\caption{Effect of the number of IoTDs.}
	\label{fig:result_no_iotds}
\end{figure}

In Fig. \ref{fig:result_no_iotds}, we vary the number of IoTDs in the network and look at the number of served devices (percentage), under the four different methods. 
First, it can be seen that with smaller devices (10), the gain was relatively high, especially for our proposed solution. The difference between the proposed solution and the nearest one is greater than 50\% all the way through. DRL+BCD manages to serve approximately 9 IoTDs when there are 10 IoTDs in the network. However, as the number of IoTDs increases, the service rate goes down. This is due to the limited wireless resources available for the UAV. Hence, when more IoTDs are added to the network, the ability of the UAV to fulfill the network needs will degrade. However, this issue can still be tackled by increasing the number of RIS elements\footnote{Fig. \ref{fig:result_no_elements} shows that the number of RIS elements has a great impact on the number of IoTDs served. However, determining the size of the RIS is another problem and is beyond the scope of this paper.}. In addition to our solution approach, the other baselines also show interesting behaviours. Intuitively, they all experience decreasing trends, however, Stationary UAV and Random Walk UAV face a sudden increase at 20 IoTDs before they start to gradually decrease. The reason behind such behaviour is that owing to the fixed wireless resources and the likelihood of IoTDs presence within the UAV vicinity, there are some resources wasted when there is little IoTDs near the UAV and RIS to serve. However, as the number of IoTDs increases, there will be more chances for the UAV to find IoTDs and serve them. Eventually, when there are more IoTDs in the network than the available resource, the percentage of served IoTDs starts to decrease and that what happens with networks of size 30, 40, and 50 IoTDs. The last two baselines, Random $\Theta^n$ and Without RIS, come in the last place with significantly poor performance, with marginal improvement for Random $\Theta^n$, due to their weak RIS tuning or non availability of the RIS. 

\subsection{Effect of Data Size}
\begin{figure}[htbp]
	\centerline{\includegraphics[scale=0.25]{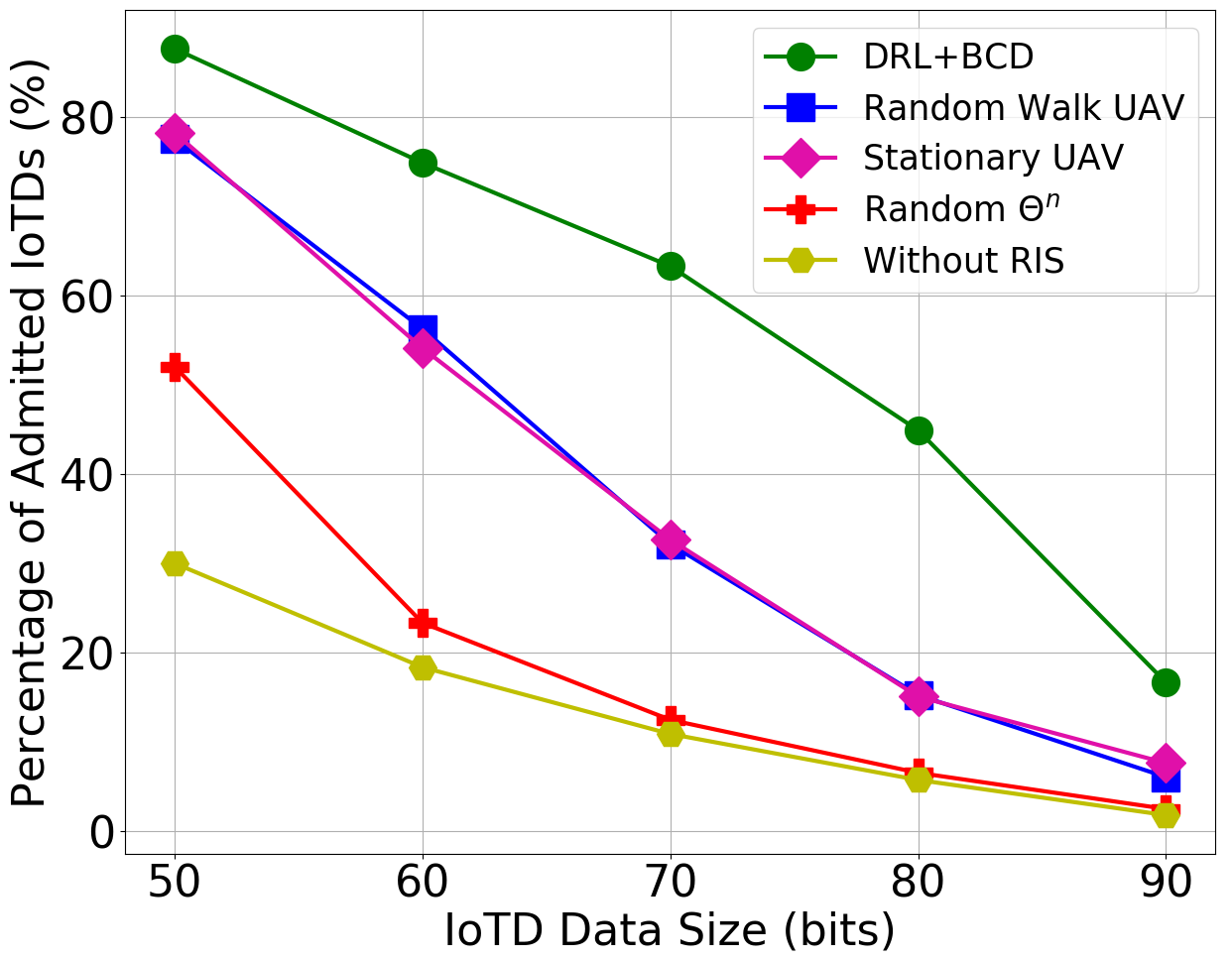}}
	\caption{Effect of the data size generated by the IoTDs.}
	\label{fig:result_data_size}
\end{figure}

In Fig. \ref{fig:result_data_size}, we examine the impacts of having small and large size of data generated by the IoTDs. The percentage of IoTDs admitted for full service is shown versus the size of data in bits. In general, the small data generated, the better service that is provided. This is due to the fact that small data can be uploaded within shorter periods, subsequently, the UAV will have more free resources to serve other requests. 

One can observe that DRL+BCD obtains the highest performance in comparison with the baselines in all scenarios. The difference between our solution approach and the other methods becomes more apparent when the data size increases. That means, our solution approach can better adapt to more complex scenarios where the UAV and RIS resource are optimized wisely. Another insight one can look at is the impact of increasing the data size (i.e., 90 bits). In such case, it becomes more difficult for the UAV to guarantee sufficient service for the IoTDs. Again, this issue can be tackled by investing more in the RIS size. Meanwhile, we can see a sharp decrease experienced by Random Walk UAV and Stationary UAV while both demonstrate better performance than Random $\Theta^n$ and without RIS (i.e. M =0). Interestingly, even with random phase-shift configuration, the RIS proves itself as a solid candidate to enhance the wireless link quality for the UAV when small size data is considered. At 50 bits, Random $\Theta^n$ is able to double the number of IoTDs admitted in comparison to the UAV alone scenario. 

\subsection{Impact of RIS Integration on UAV Energy Efficiency}
\label{sec:effect-of-ris}

\begin{figure}[htbp]
	\centerline{\includegraphics[scale=0.25]{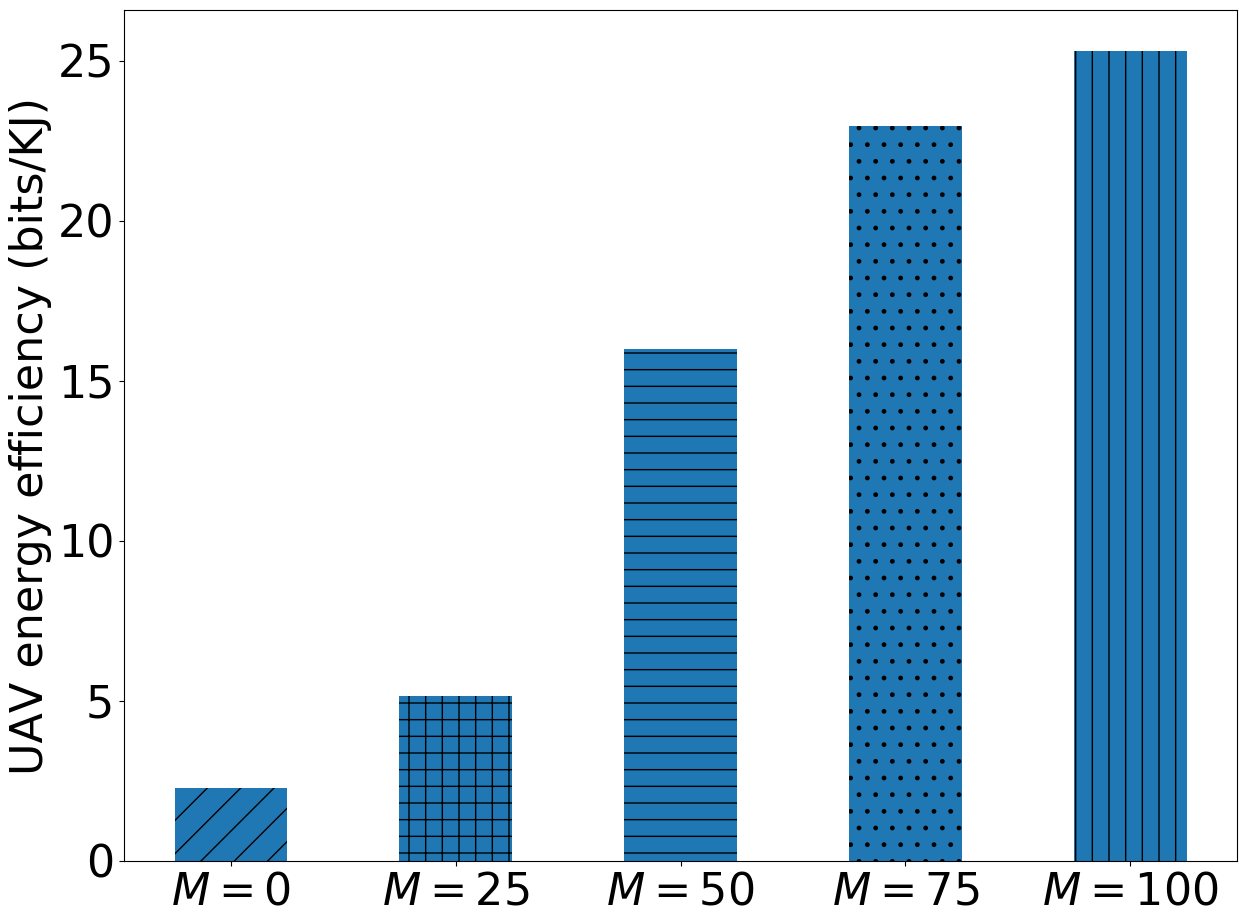}}
	\caption{UAV energy efficiency levels with various RIS sizes.}
	\label{fig:result_energy_efficiency}
\end{figure}

In Fig. \ref{fig:result_energy_efficiency}, the energy efficiency levels are obtained for various RIS sizes to understand the effect of having RIS on the UAV energy efficiency. It can be observed that without RIS, the energy consumed to upload one bit is very high. However, as the RIS size grows up, this value becomes smaller and smaller and that translates in better energy efficiency. In other words, the presence of RIS helps in two directions. First, it makes the UAV have more flexibility to plan its path such that the total amount of energy consumed at the end is minimized as per Eq \eqref{eq:uav_power_consumption}. Second, it helps to make transmission more successful by increasing the quality of the established links. This is demonstrated with RIS of 50 elements where the energy efficiency reaches above 15 bits/KJ. This significant increase keeps going up to increase by 30\% with only 25 elements added to the RIS (at $M=75$).

\section{Conclusion}
\label{sec-conclution}
In this article, we presented a new data gathering framework leveraging an RIS-UAV assisted network to collect data from sensing devices within their active times. The problem is formulated as mixed-integer non-convex programming and then, due to its complexity, is converted to MDP to be solved later via a DRL agent. The paper also proposed a BCD method to handle the RIS phase shift configuration with lower complexity and high efficiency that was proved later in the numerical results by comparing it with random coefficient settings. The superiority of our solution approach is shown through simulations against four alternative methods. Indeed, the proposed solution outperformed its counterparts by more than 50\% in many cases. Integrating RIS to the UAV communications has also shown another benefit of improving UAV energy efficiency remarkably especially when considering a large RIS.

In terms of future works, multi-UAVs can be further investigated to see how they cooperate with each other to plan their trajectory such that they avoid collisions and the ultimate objective is optimized. Moreover, the metric of Age of Information can also be applied to this context in order to ensure the freshness of the data collected from the IoTDs.

\bibliographystyle{IEEEtran}
\bibliography{IEEEabrv, reference}

\begin{thebibliography}{10}
\providecommand{\url}[1]{#1}
\csname url@samestyle\endcsname
\providecommand{\newblock}{\relax}
\providecommand{\bibinfo}[2]{#2}
\providecommand{\BIBentrySTDinterwordspacing}{\spaceskip=0pt\relax}
\providecommand{\BIBentryALTinterwordstretchfactor}{4}
\providecommand{\BIBentryALTinterwordspacing}{\spaceskip=\fontdimen2\font plus
\BIBentryALTinterwordstretchfactor\fontdimen3\font minus
  \fontdimen4\font\relax}
\providecommand{\BIBforeignlanguage}[2]{{%
\expandafter\ifx\csname l@#1\endcsname\relax
\typeout{** WARNING: IEEEtran.bst: No hyphenation pattern has been}%
\typeout{** loaded for the language `#1'. Using the pattern for}%
\typeout{** the default language instead.}%
\else
\language=\csname l@#1\endcsname
\fi
#2}}
\providecommand{\BIBdecl}{\relax}
\BIBdecl

\bibitem{chung2020design}
C.~Chung and J.~P. Jeong, ``A design of iot device configuration translator for
  intent-based iot-cloud services,'' in \emph{2020 22nd International
  Conference on Advanced Communication Technology (ICACT)}.\hskip 1em plus
  0.5em minus 0.4em\relax IEEE, 2020, pp. 52--56.

\bibitem{biason2018access}
A.~Biason, C.~Pielli, A.~Zanella, and M.~Zorzi, ``Access control for iot nodes
  with energy and fidelity constraints,'' \emph{IEEE Transactions on Wireless
  Communications}, vol.~17, no.~5, pp. 3242--3257, 2018.

\bibitem{mozaffari2016unmanned}
M.~Mozaffari \emph{et~al.}, ``Unmanned aerial vehicle with underlaid
  device-to-device communications: Performance and tradeoffs,'' \emph{{IEEE}
  Trans. Wireless Commun.}, vol.~15, no.~6, pp. 3949--3963, 2016.

\bibitem{motlagh2017uav}
N.~H. Motlagh, M.~Bagaa, and T.~Taleb, ``Uav-based iot platform: A crowd
  surveillance use case,'' \emph{IEEE Communications Magazine}, vol.~55, no.~2,
  pp. 128--134, 2017.

\bibitem{abd2018average}
M.~A. Abd-Elmagid and H.~S. Dhillon, ``Average peak age-of-information
  minimization in uav-assisted iot networks,'' \emph{IEEE Transactions on
  Vehicular Technology}, vol.~68, no.~2, pp. 2003--2008, 2018.

\bibitem{samir2020trajectory}
M.~Samir, D.~Ebrahimi, C.~Assi, S.~Sharafeddine, and A.~Ghrayeb, ``Trajectory
  planning of multiple dronecells in vehicular networks: A reinforcement
  learning approach,'' \emph{IEEE Networking Letters}, vol.~2, no.~1, pp.
  14--18, 2020.

\bibitem{samir2019uav}
M.~Samir, S.~Sharafeddine, C.~M. Assi, T.~M. Nguyen, and A.~Ghrayeb, ``Uav
  trajectory planning for data collection from time-constrained iot devices,''
  \emph{IEEE Transactions on Wireless Communications}, vol.~19, no.~1, pp.
  34--46, 2019.

\bibitem{huo2019distributed}
Y.~Huo, X.~Dong, T.~Lu, W.~Xu, and M.~Yuen, ``Distributed and multilayer uav
  networks for next-generation wireless communication and power transfer: A
  feasibility study,'' \emph{IEEE Internet of Things Journal}, vol.~6, no.~4,
  pp. 7103--7115, 2019.

\bibitem{basar2019wireless}
E.~Basar, M.~Di~Renzo, J.~De~Rosny, M.~Debbah, M.-S. Alouini, and R.~Zhang,
  ``Wireless communications through reconfigurable intelligent surfaces,''
  \emph{IEEE access}, vol.~7, pp. 116\,753--116\,773, 2019.

\bibitem{wu2019towards}
Q.~Wu and R.~Zhang, ``Towards smart and reconfigurable environment: Intelligent
  reflecting surface aided wireless network,'' \emph{{IEEE} Commun. Mag.},
  vol.~58, no.~1, pp. 106--112, 2019.

\bibitem{Elhattab}
M.~K. {Elhattab}, M.~A. {Arfaoui}, C.~{Assi}, and A.~{Ghrayeb},
  ``Reconfigurable intelligent surface assisted coordinated multipoint in
  downlink {NOMA} networks,'' \emph{IEEE Communications Letters}, vol.~25,
  no.~2, pp. 632--636, 2021.

\bibitem{cai2020joint}
Y.~Cai, Z.~Wei, R.~Li, D.~W.~K. Ng, and J.~Yuan, ``Joint trajectory and
  resource allocation design for energy-efficient secure uav communication
  systems,'' \emph{IEEE Transactions on Communications}, vol.~68, no.~7, pp.
  4536--4553, 2020.

\bibitem{lin2020dynamic}
Y.~Lin, M.~Wang, X.~Zhou, G.~Ding, and S.~Mao, ``Dynamic spectrum interaction
  of uav flight formation communication with priority: A deep reinforcement
  learning approach,'' \emph{IEEE Transactions on Cognitive Communications and
  Networking}, vol.~6, no.~3, pp. 892--903, 2020.

\bibitem{shiri2020communication}
H.~Shiri, J.~Park, and M.~Bennis, ``Communication-efficient massive uav online
  path control: Federated learning meets mean-field game theory,'' \emph{IEEE
  Transactions on Communications}, vol.~68, no.~11, pp. 6840--6857, 2020.

\bibitem{samir2020online}
M.~Samir, C.~Assi, S.~Sharafeddine, and A.~Ghrayeb, ``Online altitude control
  and scheduling policy for minimizing aoi in uav-assisted iot wireless
  networks,'' \emph{IEEE Transactions on Mobile Computing}, 2020.

\bibitem{islambouli2019optimized}
R.~Islambouli and S.~Sharafeddine, ``Optimized 3d deployment of uav-mounted
  cloudlets to support latency-sensitive services in iot networks,'' \emph{IEEE
  Access}, vol.~7, pp. 172\,860--172\,870, 2019.

\bibitem{zhan2020energy}
C.~Zhan and Y.~Zeng, ``Energy-efficient data uploading for cellular-connected
  uav systems,'' \emph{IEEE Transactions on Wireless Communications}, vol.~19,
  no.~11, pp. 7279--7292, 2020.

\bibitem{callegaro2021optimal}
D.~Callegaro and M.~Levorato, ``Optimal edge computing for
  infrastructure-assisted uav systems,'' \emph{IEEE Transactions on Vehicular
  Technology}, 2021.

\bibitem{liu2020resource}
Y.~Liu, K.~Liu, J.~Han, L.~Zhu, Z.~Xiao, and X.-G. Xia, ``Resource allocation
  and 3d placement for uav-enabled energy-efficient iot communications,''
  \emph{IEEE Internet of Things Journal}, 2020.

\bibitem{hu2019optimal}
Y.~Hu, X.~Yuan, J.~Xu, and A.~Schmeink, ``Optimal 1d trajectory design for
  uav-enabled multiuser wireless power transfer,'' \emph{IEEE Transactions on
  Communications}, vol.~67, no.~8, pp. 5674--5688, 2019.

\bibitem{yang2020performance}
L.~Yang, F.~Meng, J.~Zhang, M.~O. Hasna, and M.~Di~Renzo, ``On the performance
  of ris-assisted dual-hop uav communication systems,'' \emph{IEEE Transactions
  on Vehicular Technology}, vol.~69, no.~9, pp. 10\,385--10\,390, 2020.

\bibitem{you2021enabling}
C.~You, Z.~Kang, Y.~Zeng, and R.~Zhang, ``Enabling smart reflection in
  integrated air-ground wireless network: Irs meets uav,'' \emph{arXiv preprint
  arXiv:2103.07151}, 2021.

\bibitem{liu2020machine}
X.~Liu, Y.~Liu, and Y.~Chen, ``Machine learning empowered trajectory and
  passive beamforming design in uav-ris wireless networks,'' \emph{IEEE Journal
  on Selected Areas in Communications}, 2020.

\bibitem{wei2020sum}
Z.~Wei, Y.~Cai, Z.~Sun, D.~W.~K. Ng, J.~Yuan, M.~Zhou, and L.~Sun, ``Sum-rate
  maximization for irs-assisted uav ofdma communication systems,'' \emph{IEEE
  Transactions on Wireless Communications}, 2020.

\bibitem{shafique2020optimization}
T.~Shafique, H.~Tabassum, and E.~Hossain, ``Optimization of wireless relaying
  with flexible uav-borne reflecting surfaces,'' \emph{IEEE Transactions on
  Communications}, 2020.

\bibitem{chen2020resource}
Y.~Chen, Y.~Wang, J.~Zhang, and Z.~Li, ``Resource allocation for intelligent
  reflecting surface aided vehicular communications,'' \emph{{IEEE} Trans. Veh.
  Technol.}, 2020.

\bibitem{wang2020outage}
J.~Wang, W.~Zhang, X.~Bao, T.~Song, and C.~Pan, ``Outage analysis for
  intelligent reflecting surface assisted vehicular communication networks,''
  \emph{arXiv preprint arXiv:2004.08063}, 2020.

\bibitem{dampahalage2020intelligent}
D.~Dampahalage, K.~Manosha, N.~Rajatheva \emph{et~al.}, ``Intelligent
  reflecting surface aided vehicular communications,'' \emph{arXiv preprint
  arXiv:2011.03071}, 2020.

\bibitem{you2020channel}
C.~You, B.~Zheng, and R.~Zhang, ``Channel estimation and passive beamforming
  for intelligent reflecting surface: Discrete phase shift and progressive
  refinement,'' \emph{{IEEE} J. Sel. Areas Commun.}, vol.~38, no.~11, pp.
  2604--2620, 2020.

\bibitem{pan2020uav}
Y.~Pan, K.~Wang, C.~Pan, H.~Zhu, and J.~Wang, ``Uav-assisted and intelligent
  reflecting surfaces-supported terahertz communications,'' \emph{arXiv
  preprint arXiv:2010.14223}, 2020.

\bibitem{kouzayha2017joint}
N.~Kouzayha, Z.~Dawy, J.~G. Andrews, and H.~ElSawy, ``Joint downlink/uplink rf
  wake-up solution for iot over cellular networks,'' \emph{IEEE Transactions on
  Wireless Communications}, vol.~17, no.~3, pp. 1574--1588, 2017.

\bibitem{al2019energy}
T.~A. Al-Janabi and H.~S. Al-Raweshidy, ``An energy efficient hybrid mac
  protocol with dynamic sleep-based scheduling for high density iot networks,''
  \emph{IEEE Internet of Things Journal}, vol.~6, no.~2, pp. 2273--2287, 2019.

\bibitem{kim2020deep}
K.~H. Kim and H.-D. Kim, ``Deep sleep mode based nodemcu-enabled humidity
  sensor nodes monitoring for low-power iot,'' \emph{Transactions on Electrical
  and Electronic Materials}, vol.~21, no.~6, pp. 617--620, 2020.

\bibitem{wang2018access}
X.~Wang, X.~Chen, Z.~Li, and Y.~Chen, ``Access delay analysis and optimization
  of nb-iot based on stochastic network calculus,'' in \emph{2018 IEEE
  International Conference on Smart Internet of Things (SmartIoT)}.\hskip 1em
  plus 0.5em minus 0.4em\relax IEEE, 2018, pp. 23--28.

\bibitem{kozlowski2019energy}
A.~Koz{\l}owski and J.~Sosnowski, ``Energy efficiency trade-off between
  duty-cycling and wake-up radio techniques in iot networks,'' \emph{Wireless
  Personal Communications}, vol. 107, no.~4, pp. 1951--1971, 2019.

\bibitem{long2020reflections}
H.~Long, M.~Chen, Z.~Yang, B.~Wang, Z.~Li, X.~Yun, and M.~Shikh-Bahaei,
  ``Reflections in the sky: Joint trajectory and passive beamforming design for
  secure {UAV} networks with reconfigurable intelligent surface,'' \emph{arXiv
  preprint arXiv:2005.10559}, 2020.

\bibitem{abdullah2020hybrid}
Z.~Abdullah, G.~Chen, S.~Lambotharan, and J.~A. Chambers, ``A hybrid relay and
  intelligent reflecting surface network and its ergodic performance
  analysis,'' \emph{{IEEE} Wireless Commun. Lett.}, vol.~9, no.~10, pp.
  1653--1657, 2020.

\bibitem{samir2020optimizing}
M.~Samir, M.~Elhattab, C.~Assi, S.~Sharafeddine, and A.~Ghrayeb, ``Optimizing
  age of information through aerial reconfigurable intelligent surfaces: A deep
  reinforcement learning approach,'' \emph{arXiv preprint arXiv:2011.04817},
  2020.

\bibitem{dechene2012energy}
D.~J. Dechene and A.~Shami, ``Energy-aware resource allocation strategies for
  lte uplink with synchronous harq constraints,'' \emph{IEEE Transactions on
  Mobile Computing}, vol.~13, no.~2, pp. 422--433, 2012.

\bibitem{zeng2019energy}
Y.~Zeng, J.~Xu, and R.~Zhang, ``Energy minimization for wireless communication
  with rotary-wing {UAV},'' \emph{{IEEE} Trans. Wireless Commun.}, vol.~18,
  no.~4, pp. 2329--2345, 2019.

\bibitem{impossible}
I.~US-1, ``{IMPOSSIBLE AEROSPACE US-1 Performance Aircraft},''
  \url{https://impossible.aero/meet-us-1/}, 2021, [Online; accessed
  01-June-2021].

\bibitem{fotouhi2019survey}
A.~Fotouhi, H.~Qiang, M.~Ding, M.~Hassan, L.~G. Giordano, A.~Garcia-Rodriguez,
  and J.~Yuan, ``Survey on uav cellular communications: Practical aspects,
  standardization advancements, regulation, and security challenges,''
  \emph{IEEE Communications Surveys \& Tutorials}, vol.~21, no.~4, pp.
  3417--3442, 2019.

\bibitem{bohn2019deep}
E.~B{\o}hn, E.~M. Coates, S.~Moe, and T.~A. Johansen, ``Deep reinforcement
  learning attitude control of fixed-wing {UAVs} using proximal policy
  optimization,'' in \emph{Proc. Int. Conf. Unmanned Aircraft Syst. (ICUAS)},
  2019, pp. 523--533.

\end{thebibliography}
\end{document}